\documentclass[12pt]{article}
\usepackage[english]{babel}
\usepackage{amsmath,epsf}
\usepackage[T1]{fontenc}
\usepackage[latin1]{inputenc}
\usepackage{epsf}
\usepackage{amsfonts,amssymb}
\usepackage[usenames,dvipsnames]{color}
\newcommand{\al}{\alpha}
\newcommand{\ao}{\alpha_0}

\newcommand{\de}{\delta}
\newcommand{\De}{\Delta}
\newcommand{\vep}{\varepsilon}

\newcommand{\Ga}{\Gamma}

\newcommand{\la}{\lambda}

\newcommand{\vp}{\varphi}

\newcommand{\up}{\underline{p}}

\newcommand{\pa}{\partial}

\newcommand{\ti}[1]{\tilde{#1}}

\newcommand{\qed}{\hfill \rule {1ex}{1ex}\\ }

\newcommand{\eq}{\begin{equation}}
\newcommand{\eqe}{\end{equation}}

\newcounter{saveeqn}

\begin{document}
\message{reelletc.tex (Version 1.0): Befehle zur Darstellung |R  |N, Aufruf=
z.B. \string\bbbr}
%
%
\message{reelletc.tex (Version 1.0): Befehle zur Darstellung |R  |N, Aufruf=
z.B. \string\bbbr}
\font \smallescriptscriptfont = cmr5
\font \smallescriptfont       = cmr5 at 7pt
\font \smalletextfont         = cmr5 at 10pt
\font \tensans                = cmss10
\font \fivesans               = cmss10 at 5pt
\font \sixsans                = cmss10 at 6pt
\font \sevensans              = cmss10 at 7pt
\font \ninesans               = cmss10 at 9pt
\newfam\sansfam
\textfont\sansfam=\tensans\scriptfont\sansfam=\sevensans
\scriptscriptfont\sansfam=\fivesans
\def\sans{\fam\sansfam\tensans}
\def\bbbr{{\rm I\!R}} 
\def\bbbn{{\rm I\!N}} 
\def\bbbE{{\rm I\!E}} 
\def\bbbm{{\rm I\!M}}
\def\bbbh{{\rm I\!H}}
\def\bbbk{{\rm I\!K}}
\def\bbbd{{\rm I\!D}}
\def\bbbp{{\rm I\!P}}
\def\bbbone{{\mathchoice {\rm 1\mskip-4mu l} {\rm 1\mskip-4mu l}
{\rm 1\mskip-4.5mu l} {\rm 1\mskip-5mu l}}}
\def\bbbc{{\mathchoice {\setbox0=\hbox{$\displaystyle\rm C$}\hbox{\hbox
to0pt{\kern0.4\wd0\vrule height0.9\ht0\hss}\box0}}
{\setbox0=\hbox{$\textstyle\rm C$}\hbox{\hbox
to0pt{\kern0.4\wd0\vrule height0.9\ht0\hss}\box0}}
{\setbox0=\hbox{$\scriptstyle\rm C$}\hbox{\hbox
to0pt{\kern0.4\wd0\vrule height0.9\ht0\hss}\box0}}
{\setbox0=\hbox{$\scriptscriptstyle\rm C$}\hbox{\hbox
to0pt{\kern0.4\wd0\vrule height0.9\ht0\hss}\box0}}}}

\def\bbbe{{\mathchoice {\setbox0=\hbox{\smalletextfont e}\hbox{\raise
0.1\ht0\hbox to0pt{\kern0.4\wd0\vrule width0.3pt height0.7\ht0\hss}\box0}}
{\setbox0=\hbox{\smalletextfont e}\hbox{\raise
0.1\ht0\hbox to0pt{\kern0.4\wd0\vrule width0.3pt height0.7\ht0\hss}\box0}}
{\setbox0=\hbox{\smallescriptfont e}\hbox{\raise
0.1\ht0\hbox to0pt{\kern0.5\wd0\vrule width0.2pt height0.7\ht0\hss}\box0}}
{\setbox0=\hbox{\smallescriptscriptfont e}\hbox{\raise
0.1\ht0\hbox to0pt{\kern0.4\wd0\vrule width0.2pt height0.7\ht0\hss}\box0}}}}

\def\bbbq{{\mathchoice {\setbox0=\hbox{$\displaystyle\rm Q$}\hbox{\raise
0.15\ht0\hbox to0pt{\kern0.4\wd0\vrule height0.8\ht0\hss}\box0}}
{\setbox0=\hbox{$\textstyle\rm Q$}\hbox{\raise
0.15\ht0\hbox to0pt{\kern0.4\wd0\vrule height0.8\ht0\hss}\box0}}
{\setbox0=\hbox{$\scriptstyle\rm Q$}\hbox{\raise
0.15\ht0\hbox to0pt{\kern0.4\wd0\vrule height0.7\ht0\hss}\box0}}
{\setbox0=\hbox{$\scriptscriptstyle\rm Q$}\hbox{\raise
0.15\ht0\hbox to0pt{\kern0.4\wd0\vrule height0.7\ht0\hss}\box0}}}}

\def\bbbt{{\mathchoice {\setbox0=\hbox{$\displaystyle\rm
T$}\hbox{\hbox to0pt{\kern0.3\wd0\vrule height0.9\ht0\hss}\box0}}
{\setbox0=\hbox{$\textstyle\rm T$}\hbox{\hbox
to0pt{\kern0.3\wd0\vrule height0.9\ht0\hss}\box0}}
{\setbox0=\hbox{$\scriptstyle\rm T$}\hbox{\hbox
to0pt{\kern0.3\wd0\vrule height0.9\ht0\hss}\box0}}
{\setbox0=\hbox{$\scriptscriptstyle\rm T$}\hbox{\hbox
to0pt{\kern0.3\wd0\vrule height0.9\ht0\hss}\box0}}}}

\def\bbbs{{\mathchoice
{\setbox0=\hbox{$\displaystyle     \rm S$}\hbox{\raise0.5\ht0\hbox
to0pt{\kern0.35\wd0\vrule height0.45\ht0\hss}\hbox
to0pt{\kern0.55\wd0\vrule height0.5\ht0\hss}\box0}}
{\setbox0=\hbox{$\textstyle        \rm S$}\hbox{\raise0.5\ht0\hbox
to0pt{\kern0.35\wd0\vrule height0.45\ht0\hss}\hbox
to0pt{\kern0.55\wd0\vrule height0.5\ht0\hss}\box0}}
{\setbox0=\hbox{$\scriptstyle      \rm S$}\hbox{\raise0.5\ht0\hbox
to0pt{\kern0.35\wd0\vrule height0.45\ht0\hss}\raise0.05\ht0\hbox
to0pt{\kern0.5\wd0\vrule height0.45\ht0\hss}\box0}}
{\setbox0=\hbox{$\scriptscriptstyle\rm S$}\hbox{\raise0.5\ht0\hbox
to0pt{\kern0.4\wd0\vrule height0.45\ht0\hss}\raise0.05\ht0\hbox
to0pt{\kern0.55\wd0\vrule height0.45\ht0\hss}\box0}}}}

\def\bbbz{{\mathchoice {\hbox{$\sans\textstyle Z\kern-0.4em Z$}}
{\hbox{$\sans\textstyle Z\kern-0.4em Z$}}
{\hbox{$\sans\scriptstyle Z\kern-0.3em Z$}}
{\hbox{$\sans\scriptscriptstyle Z\kern-0.2em Z$}}}}
\noindent

\title{ Continuity of the four-point function
of massive $\vp_4^4$-theory above threshold}

\author{Christoph Kopper\footnote{\ kopper@cpht.polytechnique.fr} \\
Centre de Physique Th{\'e}orique, CNRS, UMR 7644\\
Ecole Polytechnique\\
F-91128 Palaiseau, France} 

\date{}

\maketitle

\begin{abstract}
In this paper we prove that  the four-point function
of massive $\vp_4^4$-theory is continuous as a function 
of its independent external momenta when posing the renormalization condition
for the (physical) mass on-shell. 
The proof is based on integral representations derived
 inductively from the perturbative flow equations of the  
 renormalization group. It
closes a longstanding loophole in rigorous renormalization 
theory in so far as it shows the feasibility of a physical
definition of the renormalized coupling. 

\end{abstract}

\section{Introduction }

Analyticity and regularity 
of Feynman-amplitudes in quantum field theory have been  
a long-standing subject of research, as well  for 
calculational aspects as for the mathematical structures lying
behind. After the pioneering work of Landau [Lan]
this area of research was particularly fruitful and active in the 1960ies 
 [ELOP], [Nak], [Tod].  In the 1970ies  the interest shifted somewhat away
from these questions. With the advent of QCD, 
analyticity and dispersion relations  were no
more viewed as central for the understanding of the theory of 
strong interactions. Still there has been much progress, in particular
 on the calculational side of the subject, afterwards,
progress which we are unable to review. 
See for example [tHV] where  a general analysis of the
singularity structure at one-loop level is achieved. A recent
book on the state of the art in  calculational techniques is [Smi].

A mathematically rigorous analysis of analyticity and regularity properties  
is considerably complicated by the fact that the physically
interesting theories need to be reparametrized and renormalized.
This largely destroys the simple homogeneity properties of the 
bare Feynman amplitudes. As  a consequence, analyticity studies 
were often performed on bare amplitudes, under the plausible asumption 
that the local counter terms introduced for renormalization, would not 
upset the results achieved for the bare theory. 
Historically one should note that a rigorous theory of renormalization
 was only at the disposal about a decade after Landau's paper.
Some rigorous results  taking into account renormalization
are due to Chandler [Cha], who shows with the aid of analytical
renormalization that renormalized Feynman amplitudes are holomorphic 
outside the Landau surfaces\footnote{For high order graphs these
surfaces are hard to visualize since their definition involves
the momenta (loop and external), the Feynman parameters and the
incidence and loop matrices at the same time.}, 
and that they are distributions, 
which - under certain restrictions -
are boundary values of holomorphic functions in the complexified momenta.
  
We also note that Minkowski space Green functions were much less
studied in mathematical physics after the advent of the papers of
Osterwalder and Schrader [OS] and related work which permit to conclude on the
existence of a relativistic theory once its Euclidean counter part has
been constructed and certain growth and regularity properties of its
Schwinger functions have been verified.

The procedure of perturbative renormalization, as it is 
 nowadays presented in text books, is
as follows~: One starts from a bare Lagrangian. This   Lagrangian
has to be complemented by counter terms to give meaningful results
for perturbative calculations. 
The precise values of these counter terms are fixed through
renormalization conditions, which express the free parameters
appearing  in the Lagrangian in such a way that the results
of calculations  agree with experiment. For example
the fine structure constant in QED could be fixed such that the 
cross section for Compton scattering at some fixed values 
of energy-momenta agrees with experiment.
In the theory of the massive
self-interacting scalar field to which  we will restrict in this
paper, one  has to determine correspondingly the renormalized coupling
$g\,$ by comparison with the experimental value of the 
boson-boson scattering cross section at some fixed physical
energy-momenta. This means one has to fix the value of the 
four-point function at those values of the external energy-momenta. 

But there is still a gap between this description 
and what we know~: Renormalized Feynman-amplitudes
are known to exist as distributions [Hep1], [Spe], [Zim], [EG]. 
This generally does 
not permit to prescribe their values at  given  external momenta
on imposing a renormalization condition. 
It is also known that there are regions in momentum 
space where the renormalized Feynman-amplitudes exist as analytic functions.
For the two-point function, if properly renormalized\footnote{such
  that the 1PI two-point function vanishes on the mas-shell}, 
this region is known to include the mass-shell. In fact we know the  1PI 
two-point function to be  analytic for $p^2 <4m^2\,$ [Hep2], [Stei], [EG], 
see also [KKS]. This means that the  mass
and  wave function renormalization can be performed at a physical
point, namely the physical mass. {\it For the four-point function, the 
analyticity domain does not include physical values of the momenta}
(where the external particles are on  mass-shell). Already 
at one-loop, there is a cut starting at $s=4m^2$ ($s$ being the total
energy in the centre-of-mass frame). On the other hand, knowing that
the  four-point function exists as a distribution, does not permit to
define a physical renormalized coupling, i.e. a number. A reasonable  minimal
requirement for such a definition is the continuity of the  four-point
function  in some region  above threshold  $s=4m^2\,$, i.e. in the
physical region.
It is the aim of the present paper to show that the
four-point function is a continuous function of the external momenta 
all over $\bbbr^{12}\,$ (taking into account momentum conservation
when counting the variables).   
With our methods one could go beyond, in the sense of proving 
H\"older continuity\footnote{From explicit calculations one might
suspect that the optimal value of $\eta$ should be $1/2\,$.}
of type $\eta\,,\ 0 < \eta < 1/3\,$,
w.r.t. the Lorentz invariant variables $p_i\cdot p_k\,$. 
We will also prove continuity of the two-point function in $\bbbr^{4}\,$.
Landau [Lan, ch.4], considered that the four-point function should be
continuous above threshold, and that the degree of singularity
of the Green functions
increased with the number of external lines and decreased with the
order of perturbation theory. While the first statement is for example
confirmed by [tHV], the second one which is based on counting the
number of integrations over Feynman parameters, seems to be too strong.

A first basic tool for the proof are the  flow
equations of the renormalization group 
which are presented in section 2. They permit to study
properties of Green functions in an inductive framework.
A second basic tool is  the $\al$-parametric representation
of Feynman-amplitudes [Nak] as introduced by Schwinger, which has 
led to a representation of renormalized  Feynman-amplitudes
particularly suited for the study of analyticity properties [BZ],
[IZ]. 
In section 3 we analyse integral representations 
for the Green functions w.r.t. those $\al$-parameters which
are  obtained with the aid of the flow equations
similarly as in [KKS]. Using these 
integral representations we prove continuity of the four-point 
function in section 4.

\section{The Flow Equations}

For a general and pedagogical review on the renormalization theory
based on flow equations  we refer to [M\"u], original papers
are [Pol], [KKS1].
We consider the theory of the massive self-interacting scalar field,
the Feynman-propagator  of which is given by
\eq
\frac{i}{p_0^2-\up^{\,2}-m^2+i\vep}\ .
\eqe
More precisely we will use the form
\eq
\frac{i}{p^2-m^2+i\vep(\up^{\,2}+m^2)}\ ,\quad \vep >0\ .
\label{zim}
\eqe
Using this form of the propagator  [Zim] the power counting theorem 
for renormalized Feynman diagrams also holds in Minkowski space,
in the sense that the Feynman amplitudes define 
Lorentz-invariant tempered distributions 
with a unique limit for $\vep \to 0\,$, see also [GeSch], [Spe]. 
We use the notations 
\eq
p = (p_0,\,p_1,\,p_2,\,p_3)\,,\ \,
p^2= p_0^2-\up^{\,2}\,,\ \,\up^{\,2}\,=\,p_1^2+p_2^2+p_3^2\ .
\eqe
The regularized flowing propagator for $0 \le \ao \le \al \le \infty\,$
is given by
\eq
C^{\ao,\al}(p)\,=\, 
\int_{\ao}^{\al} e^{i\alpha[p^2-m^2+i\vep(\up^{\,2}+m^2)]}
\ d\alpha \,=\, 
i\,\frac{ e^{i\ao[p^2-m^2+i\vep(\up^{\,2}+m^2)]}-
  e^{i\al[p^2-m^2+i\vep(\up^{\,2}+m^2)]}}{p^2-m^2+i\vep(\up^{\,2}+m^2)} \ .
\label{prop}
\eqe
Note that, for finite $\al\,$, this propagator is
an entire function of $p\,$. The full propagator is recovered by 
taking the regulator 
$\ao $ to $0$ and the flow parameter $\al$
to $\infty\,$. The derivative of $C^{\ao,\al}(p)$  also is
an entire function of $p\,$, it takes the simple form
\[
\dot{C}^{\al}(p)\,\equiv\,\partial_{\al}C^{\al,\ao }(p)\,=\, 
  e^{i\alpha[p^2-m^2+i\vep(\up^{\,2}+m^2)]}\ .
\] 
The theory we want to study is massive $\varphi_4^4$-theory. 
This means that we start from the {\it bare action}
at scale $\ao$
\eq
   L_{0}(\vp) = {g \over 4!}  \int_x \vp^4(x)  
   \ +  \int_x\ \{{1 \over 2}\, a_0 \, \vp^2(x) +
    {1 \over 2}\, b_0\,  
( \partial_{\mu}\vp)^2(x)  +
    {1 \over 4!}\, c_0 \,\vp^4(x)\} \ .
\label{nawi}
\eqe
\[
a_0\,,\  c_0 =O(\hbar)\,,\quad b_0 =O(\hbar^2)\ .
\]

The parameter $\hbar$ is introduced as usual to obtain a systematic
expansion in the number of loops.
From the bare action and the flowing propagator we may define 
Wilson's {\it flowing effective action} $L^{\ao,\al}$ 
by integrating out momenta in the region 
$\ao ^{-2} \le p^2 \le \al  ^{-2}\,$.
In Minkowski space it can be defined through
\eq
e^{ {i \over \hbar}[L^{\ao,\al}(\varphi)+ I^{\ao,\al}]}
~:=\, e ^{\hbar \De^{\ao,\al}}\
e^{ {i \over \hbar}L_0(\varphi)}
\label{gam}
\eqe
and can be recognized to be the generating functional of the connected
free propagator amputated Green functions (CAG) of the theory with
propagator $C^{\ao,\al}$ and bare action $L_0\,$. Here $\De^{\ao,\al}\,$ is the
functional Laplace operator
$\langle \de/\de \varphi, C^{\ao,\al}\,\de/\de \varphi\rangle\,$,
where $\langle f, \,g \rangle\,$ denotes the standard (real) scalar product.
For the multiplicative factor 
$ e^{{i \over \hbar}I^{\ao,\al}}\,$
to be well defined, we have to restrict the theory to finite volume. All
subsequent formulae are valid also in the thermodynamic limit since
they do not involve any more the vacuum functional (or partition function)
$I^{\ao,\al}\,$.

The fundamental tool for our study of the renormalization problem 
is  the functional  {\it Flow Equation} (FE) [M\"u]
\eq
\partial_{\al}\,L^{\ao,\al}\,=\, 
\frac{\hbar}{2}\,
\langle\frac{\delta}{\delta \varphi},\dot{C}^{\al }\,
\frac{\delta}{\delta \varphi}\rangle L^{\ao,\al}
\,-\,
\frac{1}{2}\, \langle \frac{\delta L^{\ao,\al} }{\delta
  \varphi} ,\dot{C}^{\al } \,
\frac{\delta L^{\ao,\al}}{\delta \varphi}\rangle\ .
\label{funcin}
\eqe
It is obtained by deriving both sides of (\ref{gam})
w.r.t. $\al\,$.
We then expand $L^{\al_0,\al}$ in moments w.r.t. $\varphi$
\[
(2\pi)^{4(n-1)}\,\,\delta_{\varphi(p_1)} \ldots \delta_{\varphi(p_n)}
L^{\ao,\al}|_{\varphi\equiv 0}
\ =
\ \delta^{(4)} (p_1+\ldots+p_{n})\, 
{\cal L}^{\ao,\al}_{n}(p_1,\ldots,p_{n})\ ,
\]
and also in a formal powers series w.r.t. $\hbar\,$ to select the loop
order $l$
\[
{\cal L}^{\ao,\al}_{n}\,=\,
\sum_{l= 0}^{\infty} \hbar^l\,{\cal L}^{\ao,\al}_{n,l}\ .
\]
From the functional FE (\ref{funcin}) 
we then obtain the perturbative
 FEs  for the n-point CAG by identifying
coefficients 
\eq
\partial_{\al}
{\cal L}^{\ao,\al}_{n,l} =
{1 \over 2}  \int\frac{d^4p}{(2\pi)^4}\ 
{\cal L}^{\ao,\al}_{n+2,l-1}(\ldots,-p,p)\ \dot{C}^{\al}(p) 
-
\sum_{l_i, n_i} 
\Biggl[
{\cal L}^{\ao,\al}_{n_1,l_1}
\,\,
\dot{C}^{\al}\,\,
{\cal L}^{\ao,\al}_{n_2,l_2}\Biggr]_{sym}\ ,
\label{fel}
\eqe
\[
l_1+l_2 =l\,, \quad
n_1 + n_2 =n+2 \ .
\]
Here $sym$ means symmetrization - i.e. summing over all
permutations 
  of  $(p_1,\ldots, p_{n})$ {\it modulo those}
which only rearrange the arguments of one factor.

The system of flow equations can be used to get control of the Green
functions. To this end one first has to specify the  boundary conditions.
At $\al = \al_0 $ they are determined through the form of the
bare action $L_0 =\ L^{\al_0,\al_0} $ (\ref{nawi}).
The free constants appearing  in (\ref{nawi}), the so-called 
relevant parameters of the theory,  are fixed by renormalization
conditions on the IR side. 
For the proof of  continuity properties of the Green functions,
it is helpful to separate the UV or renormalizabilty problem from
the large $\al$-problem, the  latter being directly related to the
proof of continuity.  We therefore impose renormalization conditions 
at some fixed positive
intermediate scale $0 < \xi < \infty\,$: 
\eq
{\cal L}^{\ao,\xi}_{2,l}(p)|_{p^2=m^2}=a_l^{\xi}\ ,\quad
\pa_{p^2}{\cal L}^{\ao,\xi}_{2,l}(p)|_{p^2=m^2}= b_l^{\xi}\ ,\quad
{\cal L}^{\ao,\xi}_{4,l}(p^r_1,\ldots,p^r_4)=c_l^{\xi}\ ,\quad l \ge 1
\label{renb}
\eqe
for suitably chosen $p^r_1,\ldots,p^r_4\,$ with
$(p^r_i)^2 =m^2\,$ and $\sum p^r_i =0\,$, i.e.
at physical values of the
external momenta\footnote{It is  
not possible to prove renormalizability on 
imposing  renormalization conditions at a physical
point without controlling the regularity of the Green functions
 at this point. This is due to the fact that  
the proof requires to perform Taylor expansions to go
away from  the renormalization point. When imposing 
conditions for {\it finite} $\xi\,$   
this poses no problem because, with our
regularization, the propagator $C^{\ao,\xi}(p)\,$, $\xi < \infty\,$,
 is analytic in $p\,$. This fact implies (as  will be seen)  
the analyticity of the regularized Green functions at finite $\xi$.}. 
Once the boundary conditions are specified, the renormalization
problem can be solved {\it inductively} 
by adopting an inductive scheme ascending  in $n+2l$ 
and for fixed $n+2l$ ascending in $l$. For this scheme to work
it is important to note that by definition there is no $0$-loop 
two-point function in $L^{\ao,\al}\,$.

To discuss analyticity and continuity 
properties it is preferable to work with one particle irreducible
(1PI) Green functions, the generating functional of which is obtained from
the one for connected  Green functions by a Legendre transform.
Starting from the generating functional of 
nonamputated connected Green functions $W^{\ao,\al}$ 
\eq
W^{\ao,\al}(J)
\;=\,i\, L^{\ao,\al}(C^{\ao,\al}\, J)\,-\,\frac12\,\langle J,C^{\ao,\al}
J\rangle
\eqe
one defines
\eq
i\,\Ga^{\ao,\al}(\phi)\,=\,[W^{\ao,\al}(J)\;-\,i\,
\langle J,\phi \rangle]_{J=J(\phi)}\, , \quad 
\phi(p)\,=\, \frac1i (2\pi)^4 \de_{J(-p)}\,W^{\ao,\al}(J)
\label{ga}
\eqe
with boundary terms 
\eq
 L_0(\vp)= L^{\ao,\ao}(\vp)\ , \quad 
\Ga _0(\phi) = L_0(\vp)|_{ \vp \equiv\phi}\ .
\label{so}
\eqe
On taking in (\ref{ga}) a derivative w.r.t. $\al\,$, 
and expressing the $\al$-derivative of $\Ga$ 
through the one  of $L\,$, using the FE  for $L$  
and reexpressing $L$ in terms of  $\Ga\,$, gives 
 the  flow equations (\ref{fe}),  (\ref{fek})
for the perturbative 1PI Green functions 
$\Gamma^{\ao,\al}_{n,l}\,$ [M\"u].

For our purpose the most convenient procedure is
to perform the Legendre transformation on the IR side only, 
i.e. w.r.t. the propagator $C^{\xi,\al}\,$, $\al \ge \xi\,$. 
By the renormalization group property we have
\[
L^{\ao,\al}(\vp)\,= \, L^{\xi,\al}(\vp)
\]
for $\ao \le \xi \le \al\,$, understanding that the boundary
value on the r.h.s. is
\[
L^{\xi,\xi}(\vp)\equiv L^{\ao,\xi}(\vp) \  .
\]
Otherwise stated, $\,L^{\xi,\xi}\,$ now takes the role of the 
bare action. In analogy with (\ref{so}) we then impose 
\eq
\Ga^{\xi,\xi}(\phi) = L^{\xi,\xi}(\vp)|_{ \vp \equiv\phi}\ .
\label{start}
\eqe 
By performing the  Legendre transformation 
w.r.t. the IR propagator $C^{\xi,\al}$ we obtain
the generating functional $\Ga^{\xi,\al}(\phi)\,$
of the connected functions, irreducible w.r.t. $\,C^{\xi,\al}\,$. 
As indicated above  we obtain the FE for these IR 1PI functions
\eq
\partial_{\al} \,\Gamma^{\xi,\al}_{n,l}(p_1,\ldots,p_{n-1}) =
\int \frac{d^4p}{(2\pi)^4}\  
{\hat \Gamma}^{\xi,\al}_{n+2,l-1}(p_1,\ldots,p_{n-1},-p,p)\ 
\dot{C}^{\al} (p)\ ,
\label{fe}
\eqe
where 
$\Ga^{\xi,\al}_{n,l}$ ($l\geq 1$) is the regularized 
connected n-point function at loop order $l\,$ in perturbation theory,
one-particle irreducible w.r.t. the IR propagator $C^{\xi,\al}\,$. 
The ${\hat \Ga}_{n,l}^{\xi,\al}$ are auxiliary functions, 
which can be expressed recursively in terms
of the $\Ga_{n,l}^{\xi,\al}$~:
\eq
{\hat \Gamma}^{\xi,\al}_{n+2,l} =
\sum_{c \ge 1}(-1)^{c+1} \sum_{l_k, n_k}\ \Bigl[ \Bigl(\,\prod_{k=1}^{c-1}  
\ \Gamma^{\xi,\al}_{n_k+2,l_k}\ {C}^{\xi,\al}(q_k) \, \Bigr)\
\Gamma^{\xi,\al}_{n_c+2,l_c}\,\Bigr]_{sym} \ \,,
\label{fek}
\eqe
\[
\sum_{k=1}^c l_k = l\ ,\quad \sum_{k=1}^c n_k =n\ .
\]
The momentum arguments $q_k$ are determined by momentum conservation.
They are given by the loop momentum $p$ plus a subsum of incoming
momenta $p_i\,$. All other momentum arguments have been suppressed.
As in (\ref{fel}) one has to symmetrize w.r.t. the external momenta
\footnote{By momentum conservation we write 
$\Gamma^{\xi,\al}_{n,l}(p_1,\ldots,p_{n-1})\,$ as a function of
$n-1\,$ momenta though it has to be noted that they are symmetric
functions of $n$ momenta, where any one of them can be expressed in terms
of the others  by momentum conservation.}.

The  CAG ${\cal L}^{\ao,\al}_{n,l}\,$
can be expressed in terms of the  $\Ga^{\xi,\al}_{n,l}\,$
by connecting them via propagators $\,C^{\xi,\al}\,$ in all possible
ways, as usual. 
One immediately realizes that an inductive scheme in the
loop order $\,l\,$ is viable for bounding
the solutions of the  1PI FE. 

The FE for 1PI Green functions (1PI w.r.t. the full propagator)
was used in 
[KKS] to obtain an integral representation  for these functions 
on successivley integrating the FE.
This representation together with results from distribution theory
[GeSch], [Spe] permits to obtain the following results,
valid also for $\ao \to  0\,$~:\\
1) The relativistic 1PI Green functions are  
 {\it  Lorentz-invariant tempered distributions}.\\
2) For external momenta $\,(p_{01},\underline{p}_1,\ldots,p_{0n},
\underline{p}_{n}) \,$
with  $|\sum_{i\in J } p_{0i} | < 2m$
 $\, \forall J \subset \{1,\ldots,
 n\}$ they agree\footnote{up to a factor of $i ^{V-1}\,$, $V$ being
   the number of vertices} with the Euclidean ones
for $\,(ip_{01},\underline{p}_1,\ldots,ip_{0n},\underline{p}_{n}) \,$ 
and are 
 {\it  smooth functions} 
in the (image of the) corresponding domain (under the Lorentz
group). For $\,|\sum_{i\in J } p_{0i} | < 2m\,$
they are analytic in each of the complex time-like
momentum variables 
$\,p_{01},\ldots,p_{0n}\,$.\\
These results imply in particular that $\,\Ga^{\ao,\infty}_{2,l}(p)\,$ 
is analytic in a neighbourhood of the mass-shell.

It is our aim to show inductively that for arbitrarily chosen 
$b_l^{\xi}\,$, $c_l^{\xi}\,$, and with $a_l^{\xi}\,$ chosen such that 
$\,\Ga^{\xi,\infty}_{2,l}(p)|_{p^2=m^2}= 0\,$,
the four-point function is a continuous function of
$p_1,\ldots,p_4\,$ (uniformly in $\ao\,$).
The same will be shown for the two-point function.
Since the renormalization conditions
at $\al=\xi$ and at  $\al=\infty$ are in one-to-one relation, it is
then evident that the four and two-point functions are continuous for 
arbitrary physical renormalization conditions respecting
$\,\Ga^{\xi,\infty}_{2,l}(p)|_{p^2=m^2}= 0\,$. 
We note in passing that $\,\Ga^{\xi,\infty}_{2,l}(p)|_{p^2=m^2}= 0\,$
implies $\,{\cal L}^{\ao,\infty}_{2,l}(p)|_{p^2=m^2}= 0\,$, since a general
contribution to  $\,{\cal L}^{\ao,\infty}_{2,l}(p)|_{p^2=m^2}\,$ is
obtained by joining together $(n+1)$ kernels 
$\,\Ga^{\xi,\infty}_{2,l_i}(p)|_{p^2=m^2}\,$ via $n$ propagators 
$\,C^{\xi,\infty}(p)|_{p^2=m^2}\,$.

The two-point function depends on $p^2$ 
only\footnote{In slightly abusive notation
we will write subsequently $\Ga_{2}(p^2)\,$ or $\Ga_{2}(p_{\vep}^2)\,$ 
instead of $\Ga_{2}(p)\,$.} . More precisely, for $\vep>0\,$ it depends on
 $p_{\vep}^2\,$ (see (\ref{vep}) below). 
Therefore we can use Schl\"omilch's interpolation formula to
decompose\footnote{For $\al < \infty\,$ the two-point 
function is an analytic function of $p^2\,$, as will be seen in the
subsequent inductive proof. For $\al =\infty\,$ it is still analytic
for $Re\, p^2 < 4 m^2\,$ and $Im\, p^2 >0\,$.}  
it as   
\eq
\Ga^{\xi,\al}_{2,l}(p_{\vep}^2)\,=\,
\Ga^{\xi,\al}_{2,l}(m^2)\,+\,
(p_{\vep}^2\,-\,m^2)\,\int_0^1 d\tau\ \pa_{p^2}\ 
\Ga^{\xi,\al}_{2,l}((1-\tau)m^2+\tau p_{\vep}^2)\ .
\label{2p}
\eqe
We want to impose
\eq
\Ga^{\xi,\infty}_{2,l}(m^2)\,=\,0
\label{2pt}
\eqe
which implies
\eq
\Ga^{\xi,\al}_{2,l}(m^2)\,=\,
\int_{\al}^{\infty} d\al'\ \pa_{\al'} \Ga^{\xi,\al'}_{2,l}(m^2)\ .
\label{2bb}
\eqe
To guarantee (\ref{2pt}), we write the two-point function 
as a solution of the FE 
\eq
\Ga^{\xi,\al}_{2,l}(p_{\vep}^2)\,=\,
\int_{\xi}^{\al}d\al_s\ \partial_{\al_s}
\Ga^{\xi,\al_s}_{2,l}(p_{\vep}^2)\,-\,
\int_{\xi}^{\infty} d\al_s \ \partial_{\al_s}
\Ga^{\xi,\al_s}_{2,l}(m^2)
\label{2bd}
\eqe 
\eq
=\ 
\int_{\xi}^{\al} d\al_s \ \partial_{\al_s}
\left( \Ga^{\xi,\al_s}_{2,l}(p_{\vep}^2)
\,-\,\Ga^{\xi,\al_s}_{2,l}(m^2)\right)
-\ \int_{\al}^{\infty} d\al_s \ \partial_{\al_s}
\Ga^{\xi,\al_s}_{2,l}(m^2)\  ,
\label{2bd2}
\eqe
where  the second term on the r.h.s. of (\ref{2bd}) 
is a constant w.r.t. $\al\,$, chosen such that (\ref{2pt}) holds. It 
will be shown to be finite in the inductive proof
so that it gives an admissible finite boundary term 
\[
a_l^{\xi} \,=\,\Ga^{\xi,\xi}_{2,l}(m^2)\,=\,
\,-\,
\int_{\xi}^{\infty} d\al_s \ \partial_{\al_s}
\Ga^{\xi,\al_s}_{2,l}(m^2)\ .
\]
In the next section we will apply the decomposition
(\ref{2bd2}), whenever there appears a two-point function
on the r.h.s. of the FE.

\section{Integral representations and large  $\al$ behaviour}

The following {\it integral representation} was  proven inductively
with the aid of the FE together with the subsequent properties 
in [KKS]\footnote{In fact this
  integral representation was proven in [KKS] for the one-particle
  irrreducible Green functions $\Ga^{\ao,\xi}_{n,l}(\vec{p})\,$.
It can be proven in the same way for the connected Green
functions starting from the FE for those. It can also be deduced
from the integral representation for the
$\Ga^{\ao,\xi}_{n,l}(\vec{p})\,$, noting that the  
${\cal L}^{\ao,\xi}_{n,l}(\vec{p})\,$ are sums of products of the
$\Ga's\,$ joined by propagators $C^{\ao,\xi}_{n,l}(\vec{p})\,$ for which we use
(\ref{prop}). The integral representation (\ref{int}) then also holds
for sums  of products of terms of the type (\ref{int}).
In [KKS] the integral representation was written  for the case of
vanishing renormalization conditions. It is easily seen to be
valid also for nonvanishing ones. One only has to be aware of the fact
that in this case the number of internal lines is no more fixed in
terms of the number of loops and of external lines since the
renormalization  constants may be of loop order $\ge 1$ themselves, a
fact which we have already taken into account in (\ref{int}),
(\ref{dar}).}. The 
statements are valid for general renormalization conditions 
at $\al =\xi\,$, that means in particular for renormalization
conditions of the form (\ref{renb}) with $\ao$-independent
(or weakly $\ao$-dependent) renormalization constants $a_l^{\xi}\,,\
b_l^{\xi}\,,\ c_l^{\xi}\,$. We have~:\\
{\it The perturbative CAG $\,{\cal L}^{\ao,\xi}_{n,l}\,$ 
can be written as finite sum of integrals of the form }
\eq
{\cal L}^{\ao,\xi}_{n,l}(\vec{p})\,=\,\sum_j 
\int_0^1 d\la_1\ldots d\la_{\sigma_{j}}\int_{\ao}^{\xi}d\xi_1
\ldots d\xi_{s_{j}}
\,G^{\xi,(j)}_{n,l}(\xi_1,\ldots,\xi_{s_{j}},\la_1,\ldots, 
\la_{\sigma_j},\vec{p})\ .
\label{int}
\eqe
Here $\vec{p}= (p_1,\ldots,p_{n-1})\,$;
$s_j\,$ is the number of internal lines  in the respective
contribution.\\
We shall set 
 $\vec{\xi}= (\xi_1,\ldots,\xi_{s_{j}})$, 
$\ \vec{\la}=(\la_1,\ldots,\la_{\sigma_{j}})$, 
$\ d\vec{\xi}= d\xi_1\ldots d\xi_{s_{j}}\,$, $\ d\vec{\la}=
d\la_1\ldots d\la_{\sigma_{j}}\,$.\\
{\it The functions $G^{\xi,(j) }_{n,l}(\vec{\xi},\vec{\la},\vec{p})$
can be written as
\eq
G^{\xi,(j)}_{n,l}(\vec{\xi},\vec{\la},\vec{p})\,=\,
V^{\xi,(j)}(\vec{\xi}) \,Q^{(j)}(\vec{\xi},\vec{\la})\,
P_{\vep,j}(\vec{p})\,
e^{i[(\vec{p},\,A_j(\vec{\xi},\vec{\la})\vec{p})_{\vep}
-m_{\vep}^2\sum_{k=1}^{s_{j}}\xi_k ]}\ .
\label{dar}
\eqe}
We denote by $(\vec{p},\,A_j(\vec{\xi},\vec{\la})\,\vec{p})_{\vep}\,$ 
a sum of scalar products  
$\sum_{k,v}(A_j)_{kv}(\vec{\xi},\vec{\la})(\,p_k\cdot p_v)_{\vep}\,$,
where 
\eq
(p_k\cdot p_v)_{\vep}=p_{0,k}\, p_{0,v} -(1-i\vep)\up_k
\,\up_v\ ,\quad  
m_{\vep}^2=(1-i\vep)m^2\ .
\label{vep}
\eqe 
{\it The matrices $A_j$ are positive-semidefinite symmetric
$(n\!-\!1)\!\times(n\!-\!1)$-matrices
which are rational functions, 
homogeneous of degree $1$ in $\vec{\xi}$ and continuous w.r.t.
$\vec{\xi},\,\vec{\la}\,$ (within the support of the integral).\\
The $V^{\xi,(j)}$'s are products of $\theta$-functions
of arguments $(\xi_i -\xi_k)\,$ 
which constrain the $\vec \xi$-integration domain. 
They stem from successively integrating the FE.\\
 The $P_{\vep,j}$ are products of  monomials in the scalar products
$(p_k\cdot p_v)_{\vep}\,$.\\
The $Q^{(j)}$ are rational functions in 
$\vec{\xi}$, $\vec{\la}$, homogeneous of degree 
$d_j\in \bbbz$ in $\vec{\xi}\,$,
and absolutely integrable for $\xi_i \to 0\,$.}

\noindent
The proof of these statements is in [KKS]. There it is also shown that
$d_j > -{s_{j}}\,$. This lower bound on 
$d_j\,$ is at the origin of the   absolute integrability of 
$G^{\xi,(j)}_{n,l}\,$ when taking $\,\ao \to 0\,$.
The $\la$-integrals stem from 
successive use of interpolation formulas, similarly as the 
$\tau$-integral in (\ref{2p}).
We do not comment further on the proof here, since the subsequent statements
on the large $\al$-behaviour of Green functions are proven with the
aid of the same techniques.\\
As a consequence of these facts one realizes that,
for $0<\ao <\xi < \infty\,$, the functions 
$\, {\cal L}^{\ao,\xi}_{n,l}(\vec{p})\,$ are analytic functions
of $\vec p\,$.\\

We now regard $\al \ge \xi\,$ with the aim to analyse the behaviour
for $\al \to \infty\,$. We call infrared lines those with propagators
$C^{\xi,\al}\,$, and ultraviolet lines those with propagators
$C^{\ao,\xi}\,$. We want to prove the following

\noindent
{\bf Proposition}~:\\
{\it We have an integral representation for 
$\, \Ga ^{\xi,\al}_{n,l}(\vec{p})\,$ 
in terms of a finite sum\footnote{there also appear contributions
which vanish for $\vep \to 0\,$ (as distributions). They are described
in the end of the proposition.}
 of integrals, of the following type~:
\eq
\Ga ^{\xi,\al}_{n,l}(\vec{p})\,=\,\sum_j 
\int_{\xi}^{\infty} d\vec\al \int d\vec\tau\int d(\vec\xi, \vec \la)\
F_j(\vec\xi, \vec \la)\ \Theta^{\al,(j)}(\vec \al)\
Q^{(j)}(\vec \xi, \vec \la,\vec \al, \vec \tau)\  P_{j,\vep}(\vec p)\ \cdot
\label{hyp}
\eqe
\[
\cdot \ 
e^{i[(\vec{p},\,A_j(\vec \xi, \vec \la,\vec \al, \vec  \tau)\vec{p})_{\vep}
\,+\,m^2 \,A_j^{(m)}(\vec \xi, \vec \la,\vec \al, \vec \tau)
\,-\,m_{\vep}^2\,\sum_{ir}^{(j)}\alpha_k] } \ 
\prod_{f=1}^{c_j} \Ga ^{\xi,\al_{i_f}}_{2,l_f}(m^2)\ ,
\quad \sum l_f < l\ .
\]
i) The factors $F_j(\vec\xi, \vec \la)\,$ are of the form 
\eq
F_j(\vec \xi, \vec \la)\,=\,
V^{\xi,(j)}(\vec \xi)\ Q^{(j)}(\vec \xi, \vec \la)
\ e ^{-i m^2_{\vep}\sum_{uv}^{(j)}
 \xi_i}\ ,
\label{fj} 
\eqe
and the properties of $\,V^{\xi,(j)}(\vec \xi)\,$, 
$Q^{(j)}(\vec \xi, \vec \la)\,$, 
as well as those of the integration variables $\vec \la,\, \vec \xi$
are listed after (\ref{dar}), (\ref{vep}).
The sum $\sum_{uv}^{(j)}   \xi_i\,$ is over the internal UV lines, 
excluding those inside the factors
$ \Ga ^{\xi,\al_{i_f}}_{2,l_f}(m^2)\,$.\\[.1cm]
ii) The matrices $A_j( \vec \xi,\vec \la,\vec \al, \vec \tau)$
are positive-semidefinite symmetric
$(n\!-\!1)\times(n\!-\!1)$-matrices. Their elements are rational
functions,  homogeneous of degree $1$, in the variables
$(\vec{\xi},\vec \alpha)$~:
\eq
A_j(\rho\, \vec \xi,\vec \la, \rho\, \vec \al, \vec \tau) =
\rho \, A_j( \vec \xi,\vec \la,\vec \al, \vec \tau)\ . 
\label{hom}
\eqe
For $\xi_i\in[0,\xi]$  and $\al_i \ge \xi\,$ 
they are continuous functions of $\vec \xi\,$ and 
smooth functions  of $\vec \al\,,\ \vec \la\,,\ \vec \tau\,$.
As functions of $\vec \al\,$ they are also rational functions.
They obey the bounds 
\eq
|A_j(\vec \xi,\vec \la, \vec \al, \vec \tau)| \ \le \ O(1)\ \sup_i \al_i
\label{lal}
\eqe
uniformly in all other parameters (within the support of the integrals).\\
In the following  we suppress the variables
$(\vec \la,\vec \tau)\,$, since they are pure spectators. 
We also  suppress  the subscript $j\,$.
The matrix elements $A_{kv}\,$ of $A\,$
admit the decomposition (suppressing also subscripts $k, v\,$)
\eq
A( \vec \xi, \vec \al) =
A_0(\vec \xi, \vec \al) + A_1( \vec \xi, \vec \al) +
A_2( \vec \xi, \vec \al)\ .
\label{deco}
\eqe
Here the functions $A_0,\  A_1,\ A_2\,$ are rational
functions,  homogeneous of degree $1$, 
and they have the same continuity and smoothness 
properties as $A$ above.  
Furthermore they have the following properties
\eq
 A_0(\vec \xi,\rho\vec\al) =  
\rho \, A_0(\vec \xi,\vec\al)\, ,\ 
  A_1(\vec \xi,\rho\vec \al)  =   A_1(\vec \xi,\vec \al)
 \, ,\ 
|\pa_{\rho}^{n}  A_2(\vec \xi,\rho \vec \al)| \le  
O(\rho ^{-1-n})\ ,
\label{asy}
\eqe
where $\rho>0\,$ and $n\in \bbbn_0\,$.
The matrix $(A_0)\,$ is also positive definite.\\
Finally $\,A_j^{(m)}(\vec \xi, \vec \la,\vec \al, \vec \tau)\,$
may be viewed as a $1\times1$-matrix with the same properties
as the  $\,A_j(\vec \xi, \vec \la,\vec \al, \vec \tau)\,$.
\\[.1cm] 
iii) The $Q^{(j)}(\vec \xi, \vec \la, \vec \al,\vec \tau)$ are rational 
functions\footnote{they may also depend on $m^2$ which we view as
  constant, however}  of  $\vec{\xi}$, $\vec{\al}$, 
which are uniformly bounded for $\xi_i \in [0,\xi]\,$. 
They admit a similar decomposition as (\ref{asy}) (with the same notation)
\[
Q(\vec \xi, \vec \al) \,=\,  Q_0(\vec \xi,\vec\al)+ 
Q_1(\vec \xi,\vec\al)+ Q_2(\vec \xi,\vec\al)\ ,
\]
\eq
 Q_0(\vec \xi,\rho\vec\al) \,=\, 
\rho^{k} \, Q_0(\vec \xi,\vec\al)\, ,\ \,
  Q_1(\vec \xi,\rho\vec \al)  \,=\,  \rho ^{k-1}  Q_1(\vec \xi,\vec \al)
 \, ,\ \,
|\pa_{\rho}^{n}  Q_2(\vec \xi,\rho \vec \al)|  \,\le\,  
O(\rho ^{k-2-n})
\label{asyq}
\eqe
for suitable $k \in -\bbbn\,$, and the $Q_i$ have the same properties
as those listed for $Q\,$.\\ 
For $\al =\sup_i \al_i\, \ge \xi\,$, 
the functions $Q(\vec \xi, \al\vec \beta)\,$,
$\al \beta_i =\al_i\,$, are uniformly continuous in $\vec \beta\,$.  
\\[.1cm]
iv) The $P_{\vep,j}$ are products of monomials in the scalar products
$(p_k\cdot p_v)_{\vep}\,$. \\[.1cm]
v) The $\tau$-parameters are integrated each over the interval
$[0,1]\,$. The sum $\sum_{ir}^{(j)}\alpha_k\,$ is over the internal IR
lines, excluding those inside the 
$ \Ga ^{\xi,\al_{i_f}}_{2,l_f}(m^2)\,$.
Assuming their number to be $s\,$, we write $\vec \al
=(\al_1,\ldots,\al_s)\,$. For $n \ge 4\,$
and for two-point functions of arbitrary momentum $p^2\,$,
the $\Theta^{\al,(j)}(\vec \al)\,$ are products of $\theta$-functions
of arguments $(\al_i -\al_k)\,$, and of one
 $\,\theta$-function $\,\theta(\al -\al_s)\,$. 
In the expression for
$\, \Ga ^{\xi,\al}_{2,l}(m^2)\,$, 
there appears one $\,\theta$-function $\,\theta(\al_s -\al)\,$ instead of
$\,\theta(\al -\al_s)\,$. 
\\[.1cm]
vi) For $n \ge 4\,$ we have the following bounds, uniformly
in $\vec \xi,\vec \al, \vec \tau$
\eq
\int_{\xi}^{\infty}d{\vec \al^{\,''}}\ |\,\Theta^{\al,(j)}(\vec \al)\
 Q^{(j)}(\vec \xi,\vec \la,\vec \al, \vec
 \tau)\ \prod_{f=1}^{c_j} \Ga ^{\xi,\al_{i_f}}_{2,l_f}(m^2)\, | \ \le \
\al^{\frac{n-4}{2}+s''-s}\ {\cal P}_{l}\log \al \ .
\label{bd}
\eqe
Here ${\cal P}_{l}\log \al\,$ denotes a polynomial \footnote{The 
coefficients of the polynomial may depend on the
parameters ($\xi, \ m,\ n,\ l\,$).} of degree $\le l$
in $\log \al\,$, and $\vec \al^{\,''}\,$ is a subset of the $\al$-parameters
$(\al_1,\ldots,\al_s)\,$ which contains $s''$ elements.\\[.1cm]
vii) The two-point functions satisfy the bound
\eq
| \,\Ga ^{\xi,\al}_{2,l}(p_{\vep}^2)|
\  \le \ O(1)\ .
\label{bd2}
\eqe
The two-point functions on mass-shell satisfy 
\eq
|\,\Ga ^{\xi,\al}_{2,l}(m^2)|\ \le \ \al^{-1} \ {\cal P}_{l}\log \al\ .
\label{bd2m}
\eqe
\noindent
For $\vep >0\,$ there also appear contributions to
$\,\Ga ^{\xi,\al}_{n,l}(\vec{p})\,$
which are of the same form as (\ref{hyp}) but which carry a factor
$\,(-i\vep m^2)^r\,, \  r\in \bbbn\,, \ r <l\,$.
For these terms the bounds (\ref{bd}, \ref{bd2}, \ref{bd2m})
are to be multiplied by $\,\al^r\,$.}
\\
 
\noindent
{\it Proof}~:\\  
The proof is based on the standard inductive scheme which goes up in
$l\,$. The statements of the Proposition then serve at the same time as 
an induction hypothesis,
and the terms appearing on the r.h.s. of
the FE (\ref{fe}), (\ref{fek}) satisfy (\ref{hyp}) -  (\ref{bd2m})
by induction.
 Starting the induction at $l=0$ is trivial since we have
$\,\Ga^{\xi,\al}_{n,0}(\vec p)\,=\, \de_{n,4}\ g\,$.
For the boundary terms at $\al =\xi\,$ (\ref{start})
the set of infrared lines with
parameters $\{\vec \al\}\,$ is  empty, as is the set
$\{\vec \tau\}\,$. For them the proposition holds
true due to (\ref{int}), (\ref{dar}) and the subsequent statements.\\[.1cm]
i) The factors $F_j$, see (\ref{fj}), 
collect together all factorized ultraviolet
contributions. Since these are not touched upon by the Gaussian 
integration in the FE, and since sums of 
products of terms of this kind still have 
the properties listed after (\ref{int}) - (\ref{vep}),
the confirmation of i) is then  obvious.\\[.1cm]
Before verifying the other items we outline some aspects of the
procedure to be followed.\\  
For $n \ge 4\,$ we will write the solutions of the  FE as 
\eq
\Ga^{\xi,\al}_{n,l}(p_1,\ldots,p_{n-1})\,=\,
\Ga^{\xi,\xi}_{n,l}(p_1,\ldots,p_{n-1})\,+\, 
\int_{\xi}^{\al}d\al_s\ \partial_{\al_s}
\Ga^{\xi,\al_s}_{n,l}(p_1,\ldots,p_{n-1})\ ,
\label{ali}
\eqe
where the second term is obtained inductively from the r.h.s. of the FE
(\ref{fe}), and the first term is obtained from (\ref{start}).\\

For $n=2\,$, once the integral representation has been proven,
 the boundary condition
 (\ref{2pt}) is implemented as follows. Starting from
(\ref{hyp}) we have  terms of the form 
\[
\int_{\xi}^{\al} d\al_s
 \int_{\xi}^{\infty} d\vec\al^{\,'} \int d(\vec\tau,\vec\xi, \vec \la)\
F(\vec\xi, \vec \la)\ \Theta^{\al_s}(\vec \al')\
Q(\vec \xi, \vec \la,\vec \al, \vec \tau)\ 
\ \cdot
\]
\[
\cdot\   P_{\vep}(p^2)
\  e^{i[p^2_{\vep}\,A(\vec \xi, \vec \la,\vec \al, \vec \tau)
\,+\,m^2\,A^{(m)}(\vec \xi, \vec \la,\vec \al, \vec \tau)
\,-\,m_{\vep}^2\,\sum_{ir} \alpha_k] } \ 
\prod_{f=1}^{c^{(i)}} \Ga ^{\xi,\al^{(i)}_{f}}_{2,l^{(i)}_f}(m^2)
\  ,\quad (\vec \al^{\,'},\al_s)=\vec \al\ .
\]
We replaced  $ \Theta^{\al}(\vec \al) \to
\Theta^{\al_s}(\vec \al')\,$ since the last integration over $\al_s$
is the new one of the induction step. 
Inserting this representation into (\ref{2bd2}) we get 
\[
\int_{\xi}^{\al} d\al_s
\int d\vec w\ {\cal F}(\vec w)\ 
\prod_{f=1}^{c^{(i)}} \Ga ^{\xi,\al^{(i)}_{f}}_{2,l^{(i)}_f}(m^2)\  
e^{i\,(m^2\,(A^{(m)}(\vec \xi, \vec \la,\vec \al, \vec \tau)\,-\,
m^2_{\vep}\sum_{ir} \alpha_k)} 
\ \cdot
\]
\[
\cdot\   \left(P( p_{\vep}^2)
\  e^{ip_{\vep}^2\,A(\vec \xi, \vec \la,\vec \al, \vec \tau)} - 
 P(m^2)\  e^{im^2\,A(\vec \xi, \vec \la,\vec \al, \vec \tau) }\right) 
\]
\eq
-\ \int_{\al}^{\infty} d\al_s 
 \int d\vec w\
\ {\cal F}(\vec w)\ 
\prod_{f=1}^{c^{(i)}} \Ga ^{\xi,\al^{(i)}_{f}}_{2,l^{(i)}_f}(m^2)\  
e^{i\,(m^2\,A^{(m)}(\vec \xi, \vec \la,\vec \al, \vec
  \tau)\,-\,m^2_{\vep}\sum_{ir} \alpha_k)}\ P(m^2)\ 
e^{i\, m^2\,A(\vec \xi, \vec \la,\vec \al, \vec \tau)}\ 
\label{m}
\eqe
with
\[
\vec w=(\vec  \xi, \vec \la,\vec \al, \vec \tau)\ ,\quad
{\cal F}(\vec \xi, \vec \la,\vec \al, \vec \tau)=
F(\vec\xi, \vec \la)\ \Theta^{\al_s}(\vec \al)\
Q(\vec \xi, \vec \la,\vec \al, \vec \tau)\ .
\] 
The difference appearing in the first term can be reexpressed 
(cf. (\ref{2pt})) as 
\eq
(p_{\vep}^2-m^2)\ 
\int_0^1 d\tau\ 
e^{i( (1-\tau)m^2+ \tau p_{\vep}^2)
\,A(\vec \xi, \vec \la,\vec \al, \vec  \tau)}
\
\{[ i A(\vec \xi, \vec \la,\vec \al, \vec \tau)\,+\, \pa_{p^2}]\,P\}
((1-\tau)m^2+ \tau p_{\vep}^2)\ .
\label{sloe}
\eqe 
Contributions from the r.h.s. of the FE
containing the first term in  (\ref{m}) are taken  together
with the propagator
\[
C^{\xi,\al}(p)\,=\, 
i\,\frac{ e^{i\xi[p_\vep^2-m_\vep^2]}-
  e^{i\al[p_\vep^2-m_\vep^2]}}{p_\vep^2-m_\vep^2} 
\]
to give the three contributions 
\eq
\left(i\, e^{i\xi(p_\vep^2-m_\vep^2)}\,-\,
  i\,e^{i\al(p_\vep^2-m_\vep^2)}\,-\,i\vep \,m^2\,C^{\xi,\al}(p)\right)\,
\int_0^1 d\tau\
e^{i ((1-\tau)m^2+ \tau p_\vep^2)\,A(\vec \xi, \vec \la,\vec \al, \vec \tau)}
 \ \ldots 
\label{case}
\eqe 
The terms $\{[ i A(\vec \xi, \vec \la,\vec \al, \vec \tau)\,+\, \pa_{p^2}]\,P\}
((1-\tau)m^2+ \tau p^2)\,$ 
have to be absorbed in the new $\,Q(\vec \xi, \vec \la,\vec \al,
\vec \tau)\,$ resp. in the new $P_\vep(p)\,$. The
term $\,e^{i(1-\tau)m^2\,A(\vec \xi, \vec \la,\vec \al, \vec
  \tau)}\,$ contributes to the terms  
 $\,e^{i m^2\,A ^{(m)}(\vec \xi, \vec \la,\vec \al, \vec
  \tau)}\,$ in the integral representation. This means that the 
$\,A ^{(m)}$-terms  are $\,A$-terms  of two-point functions, multiplied by 
factors of $(1-\tau)\,$. They therefore have the properties claimed for
the $\,A$-terms.\\

\noindent
{\it On the terms $\sim \vep^r\,$}~:\\
The bounds for the terms multiplied by  $\,\vep^r\,$, $\,r \ge 1\,$, 
generated by (iterative)
applications of (\ref{case}) and then picking the third term
\eq
i\vep \,m^2\,C^{\xi,\al}(p)\
\int_0^1 d\tau\
e^{i ((1-\tau)m^2+ \tau p_\vep^2)\,A(\vec \xi, \vec \la,\vec \al, \vec \tau)}
 \ \ldots 
\label{vepr}
\eqe
grow more rapidly in $\al\,$ than those for the other terms, by a factor
$\sim \al^r\,$. This is due to the fact that $C^{\xi,\al}(p)\,$
contributes an additional $\al\,$-integral via (\ref{prop})
not present in the other two terms from (\ref{case}).    
Therefore they will not give rise to continuous functions when
limiting them to the four-point function (see section 4).
Instead these terms can be treated exactly as in [KKS],
Corollary 12~: 
by  performing the $\al_i$-integrations, 
taking the limit $\al \, \to \, \infty\,$ and performing a homogeneity
transformation, they can be expressed as integrals in which appear 
negative powers of
quadratic forms  in the external momenta 
 with indefinite Lorentz-invariant real and positive ($\sim \vep\,$) 
imaginary part.
These quadratic forms are multiplied by absolutely integrable rational
functions and integrated  over a compact domain.
By the results of Speer [Spe], p.105,
they are then Lorentz-invariant distributions for $\,\vep \to 0\,$.
Taking into account the multiplicative factor $\,\vep^r\,$ these distributions 
thus vanish for  $\vep \to 0\,$. They are therefore not of interest for us,
and we will only consider the nonvanishing contributions from now on.
For those the regulator  $\vep>0\,$ only serves to make the Gaussian integrals 
well-defined, otherwise all bounds from now on are uniform
in $\vep\,$. {\it Therefore we will suppress from now on
the subscripts $\vep\,$ and also $j\,$ for readibility}.\\

\noindent
The integral representation (\ref{hyp}) is verified 
inductively starting from   (\ref{fe}),  (\ref{fek}).
We thus use the integral representations for the
terms $\Ga^{\xi,\al_s}_{n_k+2,l_k}\,$ on the r.h.s. of
(\ref{fek}), applying the special treatment of two-point functions
indicated previously. For all propagators $\,C^{\xi,\al_s}(q_k)\,$, 
which do not multiply a term of the type of the second term on 
the r.h.s. of (\ref{2p}), reexpressed as in (\ref{sloe}), we use 
the integral representation from (\ref{prop}). We then have to 
perform the Gaussian integral over $\,p\,$ in  (\ref{fe}) and 
afterwards the integral over $\al_s$ from $\xi$ to $\al$ to  
pass from $\pa_{\al_s}\Ga^{\xi,\al_s}_{n,l}\,$
to $\Ga^{\xi,\al}_{n,l}\,$. 
Since all contributions to the exponent of the Gaussian integral 
satisfy ii) by the induction assumption, 
and since sums over matrices with the properties from ii)
again satisfy ii), this integral has an exponent of the form 
$i \al_s p^2 +\, i\sum_{k,v=1}^{n+1} {\ti A}_{kv}\,p_k\,p_v\,$, where 
the matrix $\,({\ti A}_{kv})\,$  satisfies ii).
Here we denote $p_{n+1}=-p_n=p\,$, and $\al_s\,$ is the 
$\al$-parameter of the derived line $\dot C^{\al_s}\,$ in (\ref{fe}), 
it is the largest one in the
set of $\al$-parameters; 
$\ti A\,$ can be realized to be independent of $\al_s\,$
inductively on inspection of the FE 
\footnote{Note that $\al$-parameters larger than $\al_s\,$
only appear inside the expressions of the terms  
$\,\Ga ^{\xi,\al_s}_{2,l_f}(m^2)\,$, due to the integrals
$\int_{\al}^{\infty}\,$ in (\ref{2bd2}). These evidently do not appear
in the matrix $\ti A\,$.}.
The exponent previously given can be rearranged in a form suitable for 
integration over $p\,$ 
\eq
i \al_s p^2 +\, i\sum_{k,v=1}^{n+1} {\ti A}_{kv}\,p_k\,p_v\,=\,
i \sum_{k,v=1}^{n-1} \left[{\ti A}_{kv}\,-\,
\frac{({\ti A}_{kn}\,-\, {\ti A}_{k n+1})({\ti A}_{vn}\,-\, {\ti A}_{v
    n+1})}
{{\ti A}_{nn}+{\ti A}_{n+1n+1}-2{\ti A}_{nn+1}+\al_s}   
\right]p_k\,p_v\  +  
\label{gauss}
\eqe
\[
+\
i({\ti A}_{n+1n+1}+{\ti A}_{nn} -2{\ti A}_{nn+1}+\al_s)\
\left(p+\sum_{k=1}^{n-1}
\frac{ {\ti A}_{kn} - {\ti A}_{kn+1}}{{\ti A}_{n+1n+1}+{\ti A}_{nn}
    -2{\ti A}_{nn+1}+\al_s}\,p_k\right)^2 \ .
\]
Since ${\ti A}\,$ is positive semi-definite we have
\eq
{\ti A}_{n+1n+1}+{\ti A}_{nn}
    -2{\ti A}_{nn+1} \ge 0\ .
\label{pos}
\eqe
On performing the Gaussian integral, in the absence of polynomials
$P(\vec p)\,$, we obtain a factor of
\eq
({\ti A}_{n+1n+1}+{\ti A}_{nn} -2{\ti A}_{nn+1}+\al_s)^{-2} \le\
\al_s^{-2}\ , 
\label{gabd}
\eqe
and a new quadratic form with matrix elements
\eq
A_{kv}\ =\
{\ti A}_{kv}\,-\,
\frac{({\ti A}_{kn}\,-\, {\ti A}_{k n+1})({\ti A}_{vn}\,-\, {\ti A}_{v
    n+1})}
{{\ti A}_{n+1n+1}+{\ti A}_{nn}-2{\ti A}_{nn+1}+\al_s}
\ , \quad  1 \le k,v \le n-1\ .
\label{qf}
\eqe

We are now ready to verify the remaining items of the induction
step~:\\ 
ii) The positive semi-definiteness, homogeneity, continuity  and
smoothness properties of the
matrix $A_{kv}\,$ are verified from those of
$\ti A_{kv}\,$, for which they hold by induction, 
 with the aid of the explicit formula (\ref{qf}),
using (\ref{pos}). In particular the positive (semi-)definiteness follows
by noting that the second term on the r.h.s. of (\ref{gauss}) can be
made vanish by suitable choice of $p$, so that the first term is
nonnegative  since the l.h.s. is (on dividing by $i\,$).
Assuming by induction the decomposition (\ref{deco}) to hold for the matrix
elements of $\ti A_{kv}\,$, the contributions in the decomposition
for the matrix elements of $A_{kv}\,$ are defined as follows
\eq
A_{0,kv}(\vec \xi, \vec \al) =
{\ti A}_{0,kv}(\vec \xi, \vec \al) -
\frac{({\ti A}_{0,kn}\,-\, {\ti A}_{0,k n+1})({\ti A}_{0,vn}
\,-\, {\ti A}_{0,vn+1})}
{{\ti A}_{0,n+1n+1}+{\ti A}_{0,nn}-2{\ti A}_{0,nn+1}+\al_s}\ ,
\label{new}
\eqe
\eq
A_{1,kv}( \vec \xi, \vec \al)\, = \,
{\ti A}_{1,kv}(\vec \xi, \vec \al) \, - \, 
\frac{d_0\,e_1 +d_1\,e_0}{f_0 + \al_s} 
+\frac{d_0\,e_0\, f_1}{(f_0 + \al_s)^2}\ ,
\label{new1}
\eqe
\eq
A_{2,kv}( \vec \xi, \vec \al) \, = \,
{\ti A}_{2,kv}(\vec \xi, \vec \al)\, - \,  
\frac{d_2\,e + e_2\,d +d_1\, e_1}
{f +\al_s}
\label{new2}
\eqe
\[
\ +\ \frac{(d_0\,e_1 +d_1\, e_0)(f_1 +f_2)}
{(f_0 +  \al_s)^2}\ -\
\frac{d_0\,e_0}{f_0 + \al_s}\left\{
\frac{f_1^2 }{(f_0 +  \al_s)^2}-\frac{f_2 }{f_0 +  \al_s}\ +\
\frac{f_1\,f_2 }{(f_0 +  \al_s)^2} \right\}
\]
with the shorthands
\[
d = ({\ti A}_{kn}-{\ti A}_{kn+1})( \vec \xi, \vec \al) \, ,\ \,
e = ({\ti A}_{vn}-{\ti A}_{vn+1})( \vec \xi, \vec \al) \, ,\  \,
f = ({\ti A}_{n+1n+1}+{\ti A}_{nn}- 2{\ti A}_{nn+1})( \vec \xi, \vec \al) \, , 
\]
\eq
d_i = ({\ti A}_{i,kn}-{\ti A}_{i,kn+1})( \vec \xi, \vec \al) \ , \quad
e_i = ({\ti A}_{i,vn}-{\ti A}_{i,vn+1})( \vec \xi, \vec \al) \, ,
\label{sho}
\eqe
\[
f_i = ({\ti A}_{i,n+1n+1}+{\ti A}_{i,nn}- 2{\ti A}_{i,nn+1})( \vec
\xi, \vec \al)  \ ,\quad i \in \{0, 1,\,2\}\ .
\]
On inspection of these expressions one realizes that the properties
(\ref{asy}) are verified for the matrix elements of $A\,$ if they are
true for those of $\ti A\,$. It also follows that the $\, A_{i,kv}\,$
are rational functions, homogeneous of degree one. The positivity
of $\, A_{0}\,$ follows in the same way as that of $A\,$. Note finally
that all denominators are bounded below by $\al_s\,$, as follows
from the positivity of   $\,\ti A\,$ resp. $\,\ti A_{0}\,$.\\
Noting  that ${\ti A}\,$ is independent of $\al_s$, the bound (\ref{lal})
follows from the induction hypothesis, using  
 (\ref{qf}) and the fact that $\al_s =\sup_i \al_i\,$.\\[.1cm]
iii) The Gaussian integral is performed  with the aid of a change 
of variable $p\ \to\ \ti p =p+\sum_{k=1}^{n-1}
\frac{ {\ti A}_{kn} - {\ti A}_{kn+1}}{{\ti A}_{n+1n+1}+{\ti A}_{nn}
    -2{\ti A}_{nn+1}+ \al_s}\,p_k\,$, see (\ref{gauss}). 
Consequently the monomials from  $\, P(\vec p)$
\footnote{remember that the monomials stem initially 
from the ultraviolet boundary terms in (\ref{int})} 
which contain the variables $\pm p\,$ will lead after 
Gaussian integration to terms 
\eq
\frac{ {\ti A}_{kn} - {\ti A}_{kn+1}}{{\ti A}_{n+1n+1}+{\ti A}_{nn}
    -2{\ti A}_{nn+1}+ \al_s}\ 
\frac{ {\ti A}_{vn} - {\ti A}_{vn+1}}{{\ti A}_{n+1n+1}+{\ti A}_{nn}
    -2{\ti A}_{nn+1}+ \al_s}\ p_k \cdot p_v\ . 
\label{exm}
\eqe
Terms $\,\sim (p^2)^{n}\,$ will give rise to terms with 
exponents $-(2+n)\,$ instead of  $-2\,$ in (\ref{gabd}). 
All these contributions are rational functions 
respecting the properties claimed for 
$\, Q(\vec \xi,\vec \la,\vec \al, \vec \tau)\,$
and allowed for by the induction hypothesis.
The decomposition into $Q_0,\ Q_1,\ Q_2\,$ is performed in analogy
with (\ref{new}). For the terms from (\ref{exm}) one proceeds as in
(\ref{new})-(\ref{new2}), for those from (\ref{gabd})
we decompose using (\ref{sho}), according to 
\eq
\frac{1}{f + \al_s} 
\, = \,
\frac{1}{f_0 + \al_s}\, - \, \frac{f_1}{(f_0 + \al_s)^2}\ +\ 
\left\{
\frac{f_1(f_1+f_2) }{(f_0 +  \al_s)^2}-\frac{f_2 }{f_0 +  \al_s}\right\}
\frac{1}{f +  \al_s} 
\label{exq}
\eqe
wherefrom the dominant and subdominant scaling contributions to
$Q\,$ can be read easily on taking (\ref{exq}) to the power 2
or higher. For $\al_s \ge \xi\,$ the uniform continuity of 
$Q(\al_s\vec \beta)\,$ is evident by induction since all denominators
appearing in the new factors contributing to $Q(\al_s\vec \beta)\,$
are bounded below by $\al_s\,$.\\[.1cm]
iv) After the linear change of variables and Gaussian integration
the monomials in external momenta
obviously still have the required properties.\\[.1cm]
v) The $\tau\,$-parameters stem from the interpolation
formula (\ref{sloe}) applied to the off-shell part of the two-point
function. So there appear at most ($l-1$) $\tau$-parameters at
loop-order $l\,$. Each IR-line contributes a factor $e ^{-im^2 \al_i}\,$   
via (\ref{prop}). When performing the $\al$-integral at loop-order $l\,$ 
we integrate
\[
\int_{\xi}^{\al} d\al_s \ldots \ =\ \int_{\xi}^{\infty} d\al_s\
\theta(\al-\al_s)\ , 
\]
with the exception of the contributions stemming from terms as the 
second one in (\ref{2bd2}), where we integrate 
\[
\int_{\al}^{\infty} d\al_s \ldots \ =\ \int_{\xi}^{\infty} d\al_s\
\theta(\al_s-\al) \ .
\]
This explains the successive generation 
of  $\theta$-functions.\\[.1cm]
vi) By induction we have for the terms $\Ga^{\xi, \al_s}_{n_k+2,l_k}\,$ 
with $n_k+2 \ge 4\,$, appearing on the r.h.s. of the FE
\eq
\int_{\xi}^{\al_{s}}d{\vec \al}_k^{\,''}\ 
|\, \Theta^{\al_{s}}(\vec \al_k)\
Q_{n_k+2,l_k}(\vec \xi_k,\vec \la_k,\vec \al_k, \vec \tau_k)
\ \prod_{f=1}^{c_{j_k}} \Ga ^{\xi,\al^{(k)}_{i_f}}_{2,l_f}(m^2)\, 
\,| \ \le \
\al_{s}^{\frac{n_k+2-4}{2}+s_k''-s_k}\ {\cal P}_{l_k}\log \al_{s}\ . 
\label{pbd}
\eqe
In the presence of two-point functions ($n_k=0\,$)
we note that the contributions from the
last term in (\ref{m}) - i.e. the on-shell
 two-point functions -
are integrated from $\al^{(k)}_{i_f}$ to $\infty$ 
and can be bounded inductively by 
$(\al^{(k)}_{i_f})^{-1} {\cal  P}_{l_k}\log \al^{(k)}_{i_f}\,$,
the integrand being bounded inductively by  
$(\al^{(k)}_{i_f})^{-2} {\cal  P}_{l_k}\log \al^{(k)}_{i_f}\,$.
On the other hand terms of the form of the first one in
(\ref{2bd2}), (\ref{m})
are bounded uniformly in $\al_s\,$, using the inductive
bounds on the integrands in (\ref{2bd2}), which are
of the form $\al_s^{-2}\ {\cal  P}_{l}\log \al_{s}\,$.
If we have  a number $c'\,$ of terms of this form 
in a contribution from the r.h.s. of the FE,
we can associate with each of them  an underived propagator 
with the same momentum $q_k\,$, cf. (\ref{fek}), 
and the factor of  $\frac{1}{q_k^2-m^2}\,$
of this  accompanying propagator   compensates the corresponding
factor in  (\ref{sloe}), see (\ref{case})\footnote{The 
factor of  $\frac{1}{q_k^2-m^2}\,$ is missing in the term
$\sim \vep\,$ in (\ref{case}). This is the origin of the
additional factor of $\al\,$ in the corresponding bound,
which was mentioned after  (\ref{vepr}).} 
In total we have $c-1\,$ underived propagators in 
with $c> c'\,$\footnote{Note that there is at least one 
$\Ga ^{\xi,\al_{s}}_{n_k+2,l_k}\,$ with $n_k >0\,$ in (\ref{fek})
so that always $c> c'\,$.}.  For the remaining  $c-c'-1\,$ 
ones we use the integral representation (\ref{prop}),
which results in a contribution of  $c-c'-1\,$ - 
equal to the number of $\al_i$-integrations from (\ref{prop}) - to  the
exponent of $\al\,$ in the bound to be established,
remembering  $\al \ge \al_s \ge \al_{i}\,$.
Adding  all contributions to this exponent
resulting by induction from the bounds
on the various terms from (\ref{fe}), (\ref{fek}) - we get,
supposing that all $\al$-parameters are integrated over
\eq
\sum_{k=1}^{c-c'} \frac{n_k+2-4}{2}\ +\ (c-c'-1) -2+1\  
=\sum_{k=1}^c \frac{n_k}{2}-2
=  \frac{n-4}{2}\ .
\label{exp}
\eqe
Here the contribution $-2$ stems from the bound (\ref{gabd}) 
on the factor produced by Gauss\-ian integration,
and the contribution $+1$ corresponds to the final $\al_s$-integration
in (\ref{ali}).
For $n=4$ the $\al_s$-integral is logarithmically divergent
for $\al_s \to \infty\,$, which
leads to the appearance of a logarithm. 
Similarly   $\al_s$-integrals over the terms from (\ref{sloe}) are bounded 
logarithmically.
By induction we then arrive at
a polynomial in logarithms the degree of which is inductively  
bounded by the maximal number of divergent subintegrations, 
and therefore by the number of loops.
If some of the $\al$-parameters are not integrated over,
the above counting rules  result in the exponent from
(\ref{bd}).\\[.1cm]
vii) The bounds on the two-point functions are established 
in the same way as the previous ones. To get the improved bound
for the two-point functions
on the mass-shell, we note that due to the boundary conditions
they  are given  as integrals 
\eq
\Ga ^{\xi,\al}_{2,l}(m^2)\ =\
\int_{\al}^{\infty} d\al'\ \, \pa_{\al'}\,\Ga ^{\xi,\al'}_{2,l}(m^2)\ .
\eqe
The integrand is given by the r.h.s of the FE, and from 
(\ref{bd}) we find (by induction on lower loop orders) 
\eq
|\,\pa_{\al}\Ga ^{\xi,\al}_{2,l}(m^2)|\ \le \ \al^{-2} 
\ {\cal P}_{l}\log \al\ .
\label{bd3}
\eqe
\qed

\section{Continuity}

To verify  the continuity of  the four-point function 
$\Ga^{\xi,\al}_{4,l}(p_1,\ldots,p_4)\,$ for $\al \to \infty\,$,
we consider the integrals from (\ref{hyp}). We will
leave out the polynomials\footnote{multiplying a continuous function
  by a polynomial results again in a continuous function} in 
external momenta, which will not be touched upon, and we
suppress again  indices $j\,$
and $\vep$. For shortness we will also suppress the factors
$\,e^{im^2 \,A^{(m)}(\vec \xi, \vec \la,\vec \al, \vec \tau)}\,$
so that one should read
\eq
(\vec{p},A(\vec \xi, \vec \la,\vec \al, \vec \tau)\vec{p})
\ \to \ (\vec{p},A(\vec \xi, \vec \la,\vec \al, \vec \tau)\vec{p})\,+\,
m^2 \,A^{(m)}(\vec \xi, \vec \la,\vec \al, \vec \tau)\ .
\label{sh}
\eqe
We write as before $\vec \al = ({\vec \al}^{\,'},\al_s)\,$.
The integral contributions to $\Ga^{\xi,\al}_{4,l}(p_1,\ldots,p_4)\,$
can then be written as  
\eq
\int_{\xi}^{\al} d\al_s 
\int d{\vec \al}^{\,'} \int d\vec\tau\int d(\vec\xi, \vec \la)\
e^{i[(\vec{p},A(\vec \xi, \vec \la,\vec \al, \vec \tau)\vec{p})
-m^2\sum_{ir}\alpha_k] }\ \cdot
\label{pint}
\eqe
\[ 
\cdot\ F(\vec\xi, \vec \la)\ \Theta ^{\al_s}(\vec \al)\
Q(\vec \xi, \vec \la,\vec \al, \vec \tau)\  
\prod_{f=1}^{c} \Ga ^{\xi,\al_{i_f}}_{2,l_f}(m^2)\   .
\]
Using absolute integrability and the decomposition 
(\ref{deco}), we may rewrite (\ref{pint}) in the form 
\eq
\int_{\xi}^{\al}  d\al_s\  \al_s^{s-1}\
\int_{\xi/\al_s}^1  d\vec \beta \int d\vec\tau\ \int d(\vec\xi, \vec
\la)\ F_j(\vec\xi \, \vec \la)\  
e^{i (\vec{p},A_1(\vec \xi, \vec \la, \vec \beta, \vec \tau)\vec{p})}
\ e^{i \al_s [(\vec{p},
A_0(\vec \xi, \vec \la, \vec \beta, \vec \tau)\vec{p})
-m^2\sum_{ir}\beta_k] }\ \cdot
\label{phom}
\eqe
\[ 
\cdot\ \left(1+\sum_{r=1}^{\infty} 
\frac{[i\,(\vec{p},A_2(\vec \xi, \vec \la, \al_s \vec \beta, \al_s,  
\vec \tau)\vec{p})]^r}{r~!}\right)\
\Theta ^{\al_s}(\al_s \vec \beta)\
\ Q(\vec  \xi, \vec \la, \al_s \vec \beta, \vec \tau) \
\prod_{f=1}^{c} \Ga ^{\xi,\al_s\beta_{i_f}}_{2,l_f}(m^2)\ .
\]
Here we denote for $i \le s-1\,$, 
$\beta_i = \al_i/\al_s\,$ and  $ d\vec\al\,'= d(\al_s\,\vec
\beta)\,$. Subsequently we will write 
$A_0(\vec \xi, \vec \la, \al_s \vec \beta, \vec \tau)\,$ intead of
$A_0(\vec \xi, \vec \la, \al_s \vec \beta,1,\vec \tau)\,$  
understanding that $\beta_s=1\,$, and similarly for $Q$.
From the  Proposition we have the bound
for the four-point function integrand
\[
\int_{\xi}^{\al}  d\al_s\  \al_s^{s-1}\
\int_{\xi/\al_s}^1  d\vec \beta \
|\,  \Theta ^{\al_s}(\al_s\vec \beta)\
Q(\vec  \xi, \vec \la, \al_s \vec \beta, \vec \tau) \
\prod_{f=1}^{c} \Ga ^{\xi,\al_s\beta_{i_f}}_{2,l_f}(m^2) \, |
\ \le \
\  {\cal P}_l\log \al_s\ .
\]

In the following considerations we will leave out the factor
of $1+\sum_{r=1}^{\infty} 
\frac{[i\,(\vec{p},A_2(\vec \xi, \vec \la, \al_s \vec \beta, \al_s,  
\vec \tau)\vec{p})]^r}{r~!}$ for shortness and readibility. 
It can be easily realized that due to the large $\al_s$-fall-off 
of $\,A_2(\vec \xi, \vec \la, \al_s \vec \beta, \al_s, \vec \tau)\,$ 
we obtain the same 
large $\al_s$-bounds as those subsequently given 
on  reinserting this factor. 
The same remark holds for the $\al_s$-independent term
$e^{i (\vec{p},A_1(\vec \xi, \vec \la, \vec \beta, \vec \tau)\vec{p})}\,$.
We will also suppress the variables
$(\vec \xi, \vec \la, \vec \tau)\,$, which are  kept fixed.
We thus consider the integral
\[
\int_{\xi}^{\al}  d\al_s\ \int_{\xi/\al_s}^1  d\vec \beta \
e^{i \al_s [(\vec{p}, A_0(\vec \beta)\vec{p}) 
-m^2\sum_{ir}\beta_k] }\ 
\Theta ^{\al_s}(\al_s \vec \beta)\
\al_s^{s-1}
\ Q(\al_s \vec \beta) \
\prod_{f=1}^{c} \Ga ^{\xi,\al_s\beta_{i_f}}_{2,l_f}(m^2) 
\ .
\]
For $\,\al_s\,$ in the interval 
\[
I_{\nu} \,=\, [M^{\nu},\,M^{{\nu}+1}]\ ,\quad M>1 
\]
we split up the integration domain
${\cal I}\,$ of $\vec \beta\,$ such that
\[
{\cal D}^{({\nu})}_1(\al_s)= \{\vec  \beta \in {\cal I}\ |\ \
|(\vec p, A_0(\vec \beta)\,\vec p)
-m^2\sum_{ir}\beta_{k}| \ge M^{-\frac{2{\nu}}{3}}
\   \}\,,
\]
\[
{\cal D}^{({\nu})}_2(\al_s)= \{\vec  \beta\in {\cal I}\ |\ \
|(\vec p, A_0(\vec\beta)\,\vec p)
-m^2\sum_{ir}\beta_{k}| <
M^{-\frac{2{\nu}}{3}} \ \}\ .
\footnote{One can 
realize that the
  optimal value for splitting the domains is indeed $M^{-\frac{2{\nu}}{3}}\,$.
In this case we are left with a margin $M^{\frac{{\nu}}{3}}\,$ in both 
bounds (\ref{d0}) and (\ref{d3}) below. Therefrom it should be
possible to deduce H\"older continuity of type $\eta < 1/3\,$, 
as mentioned in the introduction.}^, 
\footnote{The 
domains depend on $\al_s\,$ through the lower bounds of the $\vec
\beta$-integrals.} 
\]
We then use partial integration to obtain\footnote{The contribution
  with the sum of $\de$-functions stems from deriving the lower bound
  of the $\beta$-integrals.}  
\[
\int_{I_{\nu}} d\al_s\ 
\int_{{\cal D}^{({\nu})}_1(\al_s)} d\vec \beta \
 e ^{i \al_s[ (\vec p, A_0(\vec \beta) \,\vec p) -
   m^2\sum_{ir}\beta_{k} ]} \ \al_s^{s-1}\
\Theta ^{\al_s}(\al_s \vec \beta)
\ Q(\vec  \xi, \vec \la, \al_s \vec \beta, \vec \tau)
\prod_{f=1}^{c} \Ga ^{\xi,\al_s\beta_{i_f}}_{2,l_f}(m^2) \ =  
\]
\[
\left[\int_{{\cal D}^{({\nu})}_1(\al_s)} d\vec\beta \ 
 \frac{e^{i\al_s[(\vec p, A_0(\vec\beta)\,\vec p)-m^2
     \sum_{ir}\beta_{k}]}}
{i[(\vec p,A_0(\vec\beta)\,\vec p)-m^2\sum_{ir}\beta_{k}]}\
\al_s^{s-1}\ \Theta ^{\al_s}(\al_s \vec \beta)
\ Q(\vec  \xi, \vec \la, \al_s \vec \beta, \vec \tau) 
\prod_{f=1}^{c} \Ga ^{\xi,\al_s\beta_{i_f}}_{2,l_f}(m^2) 
\,\right]_{M^{\nu}}^{M^{{\nu}+1}}
\]
\[
 - 
 \int_{I_{\nu}}d\al_s \int_{{\cal D}^{({\nu})}_1(\al_s)} d\vec \beta\
\frac{e^{i\al_s[(\vec p, A_0(\vec\beta)\vec
    p)-m^2\sum_{ir}\beta_{k} ]}}{i[(\vec p,
  A_0(\vec\beta)\vec
  p)-m^2\sum_{ir}\beta_{k}]}\ \cdot
\]
\eq
\cdot\ \left(\pa_{\al_s}\  -\ 
\frac{\xi}{\al_s^2}\sum_{i=1}^{s-1} \de(\beta_i -\frac{\xi}{\al_s})\right)
\ \al_s^{s-1}\ \Theta ^{\al_s}(\al_s \vec \beta)\
\ Q(\vec  \xi, \vec \la, \al_s \vec \beta, \vec \tau)\
\prod_{f=1}^{c} \Ga ^{\xi,\al_s\beta_{i_f}}_{2,l_f}(m^2)  \ . 
\label{49}
\eqe
By the Proposition 
each of the three terms on the r.h.s. 
of (\ref{49}) is suppressed by one power of
$\al_s\,$ or $M^{\nu}\,$ as compared to the original bound
on the four-point function, without counting the denominator. 
For the first term, (\ref{bd}) shows that 
suppression of the $\al_s$-integration
leads to this gain. 
Furthermore application of the derivative
$\pa_{\al_s}\,$ results in such a gain when applying it to the
$\theta$-function $\Theta ^{\al_s}(\al_s \vec \beta)\,$, and also   
when applying it to $\al_s^{s-1}\ 
Q(\vec  \xi, \vec \la, \al_s \vec \beta, \vec \tau)\,$ by the
established homogeneity properties of $Q(\vec  \xi, \vec \la, \al_s
\vec \beta, \vec \tau)\,$. Finally  
$\pa_{\al_s}\ \Ga ^{\xi,\al_s\beta_{i_f}}_{2,l_f}(m^2)\,$ is bounded 
by $\al_{s}^{-1}\ \al_{i_f}^{-1}\ {\cal P}_{l_f}\log \al_{i_f}\,$ 
inductively from the r.h.s. of the FE, using also the chain rule. 
The terms involving the $\de$-functions give contributions suppressed
by two powers of $\al_s\,$.\\
The r.h.s. 
of (\ref{49})  can therefore  be bounded by
\eq
 M^{\frac{2{\nu}}{3}}\ \cdot\ M^{-{\nu}}\ {\cal P}_{l-1}\log M^{\nu}  
 \ \le\   M^{-\frac{{\nu}}{3}}\ \cdot\ 
{\cal P}_l\log M^{\nu}\ . 
\label{d0}
\eqe
Summing over ${\nu}\in \bbbn\,$ we obtain
a bound $O(1)\,$, i.e. a bound uniform in $\al\,$.

In the region ${\cal D}^{({\nu})}_2\,$ we analyse further
the term $\ (\vec p, A_0(\vec\al)\,\vec p)- m^2\sum_{ir}\al_{k}\,$.
On inspection of (\ref{new}), remembering (\ref{sh}), 
the dependence of this expression on $\al_s\,$ can be written as
\eq
\sum_{k,v}A_{0,kv}(\vec \xi, \vec \al)\,p_k\,p_v 
\,+\, m^2\, A^{(m)}_{0}(\vec \xi, \vec \al^{\, '})
\,-\, m^2\sum_{ir} \al_k\ =\
-m^2\,\left(d\,+\,\frac{a}{b+\al_s}+\al_s\right)\ ,
\label{x}
\eqe
where
\[
a= \sum_{k,v}({\ti A}_{0,kn}\,-\, {\ti A}_{0,k n+1})({\ti A}_{0,vn}
\,-\, {\ti A}_{0,vn+1})\ \frac{p_k\,p_v}{m^2}
\]
\[
d=-  \sum_{k,v} {\ti A}_{0,kv}\ \frac{p_k\,p_v}{m^2}
\,-\,A^{(m)}_{0} \,+\,\sum_{k=1}^{s-1}
\al_k\ ,\quad
b={\ti A}_{0,n+1n+1}+{\ti A}_{0,nn}-2{\ti A}_{0,nn+1} \ge 0\ .
\]
Introducing for shortness the variable 
$x=\al_s+b \ge \al_s \ge M^{\nu}\,$, analysis of the function
\[
f(x)\,=\,\frac{a}{x}+x+d'\ ,\quad d'=d-b\ ,
\]
shows that the measure $\mu(\cal C_{\nu})\,$ 
of the set $\cal C_{\nu}\,$ of points $x\,$ such that 
$|\,f(x)\,|\ \le\ M \cdot\ M^{\nu/3} \,$
 inside $\,I_{\nu}+b\,$
satisfies\footnote{this condition on
  $\al_s$ is necessary for ${\cal D}^{({\nu})}_2\,$ to be nonempty.}$^,$
\footnote{In fact  $\cal C_{\nu}\,$  is a set of at most two
  intervals, and the constant $O(1)\,$ can be taken as $2\sqrt 2M\,$,
the bound for this choice being saturated
 if $a +x^2 +d' x\,$ has 2 zeroes at distance 
$\,2\sqrt M \,M^{\frac{2\nu}{3}}\,$ inside  $I_{\nu}+b\,$.}
uniformly in $\,a,\ d',\ b \ge 0\,$
\[
\mu({\cal C}_{\nu})\ \le \  O(1)\ M^{2\nu/3}\ .
\]
From this we obtain  
\[
|\,\int_{I_{\nu}} d\al_s 
\int_{{\cal D}^{({\nu})}_2(\al_s)} d\vec \beta\ 
\ e ^{i \al_s[ (\vec p, A_0(\vec \beta) \,\vec p) 
- m^2\sum_{ir}\beta_{k}]}\ 
\al_s^{s-1}\ \Theta ^{\al_s}(\al_s \vec \beta)\
\ Q(\vec  \xi, \vec \la, \al_s \vec \beta, \vec
\tau)\prod_{f=1}^{c} \Ga ^{\xi,\al_s\beta_{i_f}}_{2,l_f}(m^2) \,   |
\]
\[
\le\ 
\int_{I_{\nu}} d\al_s \int_{{\cal D}^{({\nu})}_2(M^{\nu+1})} d{\vec \beta}\ \
|\,\Theta ^{\al_s}(\al_s \vec \beta)\ \al_s^{s-1}\
\ Q(\vec  \xi, \vec \la, \al_s \vec \beta, \vec \tau)\
\prod_{f=1}^{c} \Ga ^{\xi,\al_{i_f}}_{2,l_f}(m^2) \,   |
\]
\[
\le\ 
[\sup_{{\vec \beta} \in {\cal D}^{({\nu})}_2(M^{\nu+1})}\, \mu({\cal C}_{\nu})]\
\int_{{\cal D}^{({\nu})}_2(M^{\nu+1})}d{\vec \beta}\ \,
 \sup_{\al_s \in I_{\nu}}
|\,\Theta ^{\al_s}(\al_s \vec \beta)\ \al_s^{s-1}\
\ Q(\vec  \xi, \vec \la, \al_s\vec \beta, \vec \tau)\
\prod_{f=1}^{c} \Ga ^{\xi,\al_{i_f}}_{2,l_f}(m^2) \,   |
\]
\eq
\le\  
O(1)\  (\frac{M^{\nu+1}}{M^{\nu}})^{s-1}\ 
M^{2\nu/3}\  M^{-\nu}\ {\cal P}_l\log M^{\nu+1}\ ,
\label{d3}
\eqe
where in the last bound we used (\ref{bd}) with $s''-s=-1\,$,
 (\ref{bd2m}) and the scaling properties of $\,Q\,$. 
The factor of $\, (\frac{M^{\nu+1}}{M^{\nu}})^{s-1}\,$
is independent of $\nu\,$ and can thus be absorbed in $\,O(1)\,$
(remember that our constants may depend on $\, l\,$ and that
$\, s \le 2l\,$ for the four-point function).
From this expression we again deduce a bound uniform in $\al\,$
on summing over ${\nu}\in \bbbn\,$.\\ 
The continuity properties of $A$ and $Q$ and the compactness of
the remaining variables then give, on summing both bounds  (\ref{d0}), 
(\ref{d3}) over $\nu$
\[
\bigl|\,\int_{\xi}^{\infty} d\al_s 
\int d\vec\al' \int d\vec\tau\int d(\vec\xi, \vec \la)\
e^{i[(\vec{p},A(\vec \xi, \vec \la,\vec \al, \vec \tau)\vec{p})
-m^2\sum_{ir}\alpha_k] }\ \cdot
\]
\[ 
\cdot
\ F(\vec\xi, \vec \la)\ \Theta ^{\al_s}(\vec \al)\
\Theta ^{\al_s}(\al_s \vec \beta)\
Q(\vec \xi, \vec \la,\vec \al, \vec \tau)\  
\prod_{f=1}^{c} \Ga ^{\xi,\al_{i_f}}_{2,l_f}(m^2) \,\bigr| \ <\ 
\infty .
\]

From this uniform bound in $\al\,$ we   
easily deduce the continuity of the four-point function.
Since (\ref{d0})\footnote{The expressions appearing in the
integrands from  (\ref{49}) are not uniformly bounded in 
$\vec p \in \bbbr^{12}$, but parameter values
for which the denominators appearing in those expressions fall (in
modulus) below $M^{-\frac{2\nu}{3}}\,$ do not belong to 
${\cal D}^{({\nu})}_1\,$.}, 
(\ref{d3}) hold uniformly in $\vec p \in \bbbr^{12}$, 
we can  choose $\nu_0\,\in \bbbn\,$ for  $\vep >0\,$ such that
$ \forall\ \vec p \in \bbbr^{12}\ $
\[
\sum_{\nu \ge \nu_0} 
|\int_{I_{\nu}} d\al_s\ 
\int d{\vec \al}^{\,'}
\ e ^{i[ (\vec p, A(\vec \al) \,\vec p) 
- m^2\sum_{ir}\al_{k}]}\
\Theta ^{\al_s}(\vec \al)\
\ Q(\vec  \xi, \vec \la, \vec \al,\vec \tau)\ 
\prod_{f=1}^{c} \Ga ^{\xi,\al_{i_f}}_{2,l_f}(m^2) \,|
\ \le \ \vep/3 \ .
\]
Calling $\Ga_j(\vec p) \,$ the contribution to the four-point
function  corresponding to the previous integral we 
can therefore split
\[
\Ga_j(\vec p)-\Ga_j(\vec p^{\,'})=
\Ga_j(\vec p)-\Ga ^{(< \nu_0)}_j(\vec p) \ +\
\Ga ^{(< \nu_0)}_j(\vec p)-\Ga ^{(< \nu_0)}_j(\vec p^{\,'}) \ +\
\Ga ^{(< \nu_0)}_j(\vec p^{\,'})-\Ga_j(\vec p^{\,'})\ .
\]
The first and last terms are then bounded in modulus
by $\vep/3\,$, and since $\Ga ^{(< \nu_0)}_j(\vec p)$ is an 
analytic function of $\vec p\,$, the second one is
 bounded by $\vep/3\,$, if we choose 
$\,|\vec p-\vec p^{\,'}|\,$ sufficiently small.

It is obvious from the present proof,  that
the two-point function is  also continuous in 
the variable $p^2\,$.  Since $\Ga^{\xi,\al}_{2,l}\,$ is uniformly bounded in
$\al\,$ by the previous section, its continuity follows
without taking into account the oscillating exponential.
With the same methods as used for the four-point function,
one can show that the two-point function is (H\"older) continuously 
differentiable in the variable $p^2\,$. 
We do not further elaborate on this since analyticity of the
 two-point function up to $p^2= 4m^2\,$ is well-known anyway.

Finally continuity of the IR-1PI four-point function implies 
also the continuity
connected (amputated) four-point function $\,{\cal L}_{4,l}^{0,\infty}\,$. 
This follows from the fact that in (symmetric) 
$\vp_4^4\,$-theory the only 1PI kernels appearing in the decomposition
of the connected  four-point function are the 1PI two-point functions and
one four-point function. For our renormalization conditions  the  IR-1PI   
two-point functions vanish on mass-shell and can be expanded 
around it by analyticity. The factors of $(p^2-m^2)\,$ coming from
this expansion cancel the denominators of the IR propagators 
joined to  the  IR-1PI two-point functions, so that after this cancellation
the connected  four-point function appears as a product of continuous
functions, which is then continuous itself.

To resume we have proven~: {\it The four-point function of $\vp_4^4\,$-theory
can be represented as a continuous function all over momentum
space.} Since it is known to be a Lorentz-invariant tempered 
distribution this function is then necessarily Lorentz-invariant too.

\newpage

\noindent {\bf Acknowledgement}:\\
 The author is indebted to Jacques Bros for instruction on analyticity
 properties of the four-point function and  to Xavier
Lacroze for numerous discussions.\\
  
\noindent {\bf References}:

\begin{itemize} 

\item[[BZ]$\!\!$] M. Berg\`ere and J.B. Zuber, Renormalization of Feynman
  Amplitudes and parametric integral representation, 
Commun. Math. Phys.  {\bf 35}, 113-140 (1974). 

\item[[Cha]$\!\!$] C. Chandler, Some physical region mass shell
  properties of renormalized Feynman Integrals, Commun. Math. Phys.
  {\bf 19}, 169-188 (1970).

\item[[EG]$\!\!$] H. Epstein and V. Glaser, The Role of Locality in
  Perturbation Theory, Ann. Inst. Poincar{\'e} {\bf XIX}, 211-295
  (1973), and~: Adiabatic Limit in Perturbation Theory, Erice
  Adavanced study Institute 1975; G. Velo and A. Wightman (eds.),
D. Reidel Publishing Company Utrecht (1976).  

\item [[ELOP]$\!\!$] R.J. Eden, P.V. Landshoff, D.I. Olive  and
  J.C. Polkinghorne, The Analytic S-Matrix, Cambridge University
  Press, Cambridge (1966).

\item[[GeSch]$\!\!$] I.M. Gelfand and G.E. Schilow, Generalized Functions
(Distributions), Academic Press, New York (1964).

\item[[Hep1]$\!\!$] K. Hepp, Proof of the Bogoliubov-Parasiuk Theorem
  on Renormalization, Commun. Math. Phys. {\bf 2}, 301-326 (1966), and~:
Th\'eorie de la Renormalisation, Lecture
  Notes in Physics 2, Springer Verlag, Heidelberg (1969). 

\item[[Hep2]$\!\!$] K. Hepp, Renormalization Theory, in: Statistical
  Mechanics and Quantum Field Theory, 429-500; C. de Witt and R. Stora
eds., Gordon and Breach, New York (1971). 

\item[[IZ]$\!\!$] C. Itzykson and J.B. Zuber,  Quantum Field Theory, Mc Graw
Hill, New York (1980).

\item[[KKS]$\!\!$]  G. Keller, Ch. Kopper and C. Schophaus, 
Perturbative Renormalization with Flow 
Equations in Minkowski Space, Helv. Phys. Acta {\bf 70}, 247-274 (1997). 

\item[[KKS1]$\!\!$]  G. Keller, Ch. Kopper and  M. Salmhofer,
Perturbative Renormalization and Effective Lagrangians
in $\phi^4_4$, Helv. Phys. Acta {\bf 65}, 33-52 (1992).

\item[[M\"u]$\!\!$]  V.F. M\"uller,  
Perturbative Renormalization by Flow 
Equations, Rev. Math. Phys. {\bf 15}, 491-558 (2003). 

\item[[Lan]$\!\!$]  L.D. Landau, On analytic properties of vertex
  parts in quantum field theory,
Nucl. Phys. {\bf 13}, 181-192 (1959). 

\item[[Nak]$\!\!$] N. Nakanishi, Graph Theory and Feynman Integrals,
Gordon and Breach, New York (1970).

\item[[OS]$\!\!$] K. Osterwalder and R. Schrader, Axioms for Euclidean
  Green's functions I, Commun. Math. Phys.  {\bf 31}, 83-112 (1973),
Axioms for Euclidean Green's functions II, 
Commun. Math. Phys.  {\bf 42}, 281-305 (1975).

\item[[Pol]$\!\!$] J. Polchinski, Renormalization and Effective
  Lagrangians, Nucl. Phys. {\bf B231}, 269-295 (1984).

\item[[Smi]$\!\!$] V.A. Smirnov, Feynman Integral Calculus,
Springer Verlag, Heidelberg (2006).

\item[[Spe]$\!\!$] E. Speer, Generalized Feynman Amplitudes, 
Annals of Math. Studies No. 62, Princeton Univ. Press, Princeton (1969).

\item[[Stei$\!\!$]] O. Steinmann, Perturbation Expansions in  Axiomatic
  Field Theory, Lecture Notes in Physics 11,  Springer Verlag,
  Heidelberg etc. (1971). 

\item[[tHV]$\!\!$]  G. 't Hooft and M. Veltman, Scalar One-Loop
  Integrals, Nucl. Phys. {\bf B513}, 365-401 (1979). 

\item[[Tod]$\!\!$] I.T. Todorov, Analyticity Properties of Feynman Diagrams
in Quantum Field Theory, Pergamon Press, Oxford (1971).

\item[[Zim]$\!\!$] W. Zimmermann,  The power counting theorem for
  Minkowski metric, Commun. Math. Phys.  {\bf 11}, 1-8 (1968), 
and~: Convergence of Bogoliubov's method of renormalization in momentum space,
   ibid. {\bf 15}, 208-234 (1969).

\end{itemize}

\end{document}